\documentclass[aps,prd,preprint,showpacs,groupedaddress,nofootinbib]{revtex4}
\usepackage{amsmath,amssymb,amsbsy,color}

\newcommand{\nn}{\nonumber}
\def\dfrac#1#2{\displaystyle\frac{#1}{#2}}
\newcommand{\ovl}[1]{\overline{#1}}

\newcommand{\pslash}{p\kern-1ex /}
\newcommand{\lslash}{l\kern-1ex /}
\newcommand{\sslash}{s\kern-1ex /}
\newcommand{\Dslash}{{\cal D}\kern-1.5ex /}

\newcommand{\beqa}{\begin{eqnarray}}
\newcommand{\eeqa}{\end{eqnarray}}

\newcommand{\be}{\begin{equation}}
\newcommand{\ee}{\end{equation}}
\newcommand{\bea}{\begin{eqnarray}}
\newcommand{\eea}{\end{eqnarray}}
\newcommand{\ba}{\begin{array}}
\newcommand{\ea}{\end{array}}

\newcommand{\pref}[1]{(\ref{#1})}

\begin{document}

\preprint{UTHEP-493}

\title{
Twisted-mass QCD, O(a) improvement and Wilson chiral perturbation theory 
}
\author{Sinya Aoki$^{1,2}$ and Oliver B\"ar$^{1}$}
\affiliation{
$^1$Graduate School of Pure and Applied Sciences, University of Tsukuba, Tsukuba 305-8571, Japan \\
$^2$Riken BNL Research Center, Brookhaven National Laboratory, Upton, 11973, USA
}

\date{\today}
%
\begin{abstract}
%
We point out a caveat in the proof for automatic O(a) improvement in twisted mass lattice QCD at maximal twist angle. With the definition for the twist angle previously given by Frezzotti and Rossi, automatic O(a) improvement can fail unless the quark mass satisfies  $m_{q}\gg a^{2} \Lambda_{\rm QCD}^{3}$. We propose a different definition for the twist angle which does not require a restriction on the quark mass for automatic O(a) improvement. In order to illustrate explicitly automatic O(a) improvement we compute the pion mass in the corresponding chiral effective theory. We consider different definitions for maximal twist and show explicitly the absence or presence of the leading O(a) effect, depending on the size of the quark mass. 
\end{abstract}

\maketitle

%
\section{Introduction}
%
Recently more and more evidence has been accumulated that the twisted mass formulation of lattice QCD (tmLQCD) \cite{Frezzotti:2000nk,Frezzotti:2001ea} with Wilson fermions has significant advantages compared to its untwisted counterpart (for reviews on the subject see Refs.\ \cite{Frezzotti:2002iv,FrezzottiFLab}). 
The presence of a non-zero twisted mass term protects the Dirac operator against very small eigenvalues and consequently solves the problem of ``exceptional configurations'' in the quenched approximation \cite{AG1,AG2}. The absence of these small eigenvalues is also very beneficial in unquenched simulations \cite{AG3}. Recent results indicate that unquenched simulations with twisted mass Wilson fermions are comparable in numerical cost to unquenched simulations with staggered fermions \cite{KennedyFLab}. Moreover, a twisted mass term simplifies the renormalization of matrix elements of certain local operators such as the isotriplet axial current and the isosinglet scalar density. Finally, it has been shown in Refs.\ \cite{Frezzotti:2003ni,Frezzotti:2004wz} that hadronic masses and certain matrix elements are automatically O(a) improved at maximal twist. 

The automatic O(a) improvement is quite remarkable since it does not require the computation of any improvement coefficients in the standard improvement program laid out by Symanzik \cite{'tHooft:1980xb,Symanzik:1983dc,Symanzik:1983gh}. This is a significant advantage taking into account that a non-perturbative determination of all improvement coefficients can be quite demanding.

In this paper we point out a caveat in the proof of automatic O(a) improvement given in Ref.\ \cite{Frezzotti:2003ni}. This caveat has its origin in the way the twist angle is defined. With the definition in Ref.\ \cite{Frezzotti:2003ni} one can show that, under certain conditions,  automatic O(a) improvement is only guaranteed if the quark mass satisfies the condition $m_{q}\gg a^{2} \Lambda_{\rm QCD}^{3}$. 
Many current lattice simulations, in particular unquenched simulations, probably do not satisfy this inequality well enough and automatic O(a) improvement might be lost. 

The restriction due the condition $m_{q}\gg a^{2} \Lambda_{\rm QCD}^{3}$, however, is not a fundamental limitation. In fact, the restriction is entirely due to the way the twist angle is defined. In this paper we propose an alternative definition for the twist angle, and with this definition automatic O(a) improvement at maximal twist holds without any restriction on $m_{q}$. 

The differences between the different definitions for the twist angle and its consequences for automatic O(a) improvement can be explicitly demonstrated using Wilson Chiral Perturbation theory (WChPT), i.e.\ the chiral effective theory for lattice QCD with Wilson fermions \cite{WCPT1,Aoki:2003yv,Bar:2003mh,Bar:2003xq} (for a review see Ref.\ \cite{BaerFLab}).
As an example we compute the pion mass including the leading lattice spacing contributions in this effective theory.  We explicitly show that the pion mass is O(a) improved as long as the inequality $m_{q}\gg a^{2} \Lambda_{\rm QCD}^{3}$ holds. However, uncanceled O(a) cut-off effects are present if this bound is violated and the definition in Ref.\ \cite{Frezzotti:2003ni} for maximal twist is used. On the other hand, if we define maximal twist employing our alternative definition, this O(a) contribution is absent for any value of the quark mass. 

This paper is organized as follows. In section 2 we briefly repeat the argument in Ref.\ \cite{Frezzotti:2003ni} for automatic O(a) improvement at maximal twist angle and point out where it can fail. We also propose an alternative definition for the twist angle, which guarantees automatic O(a) improvement irrespective of the size of $m_{q}$. In section 3  we set up WChPT for tmLQCD and use it in section 4 to discuss O(a) improvement of the pion mass for different definitions of maximal twist. Some final remarks are made in section 5.
%
\section{Critical quark mass and twist angle}
%
\subsection{Automatic O(a) improvement at maximal twist}
%
First we briefly repeat the argument given in Ref.\cite{Frezzotti:2003ni} for automatic O(a) improvement at maximal twist angle. For convenience we follow the notation introduced in this reference.

The fermion mass term of tmLQCD with Wilson fermions is defined as
\beqa
\bar\psi_{\rm ph}(x)\left[
\left( -a\frac{r}{2}\sum_\mu\nabla_\mu^\star\nabla_\mu + M_{\rm cr}(r)\right)
\exp(-i\omega\gamma_5\tau_3) + m_q\right]
\psi_{\rm ph}(x) 
\label{eq:twQCD}
\eeqa
in the so-called physical basis, while it becomes
\beqa
\bar\psi(x)\left[
\left( -a\frac{r}{2}\sum_\mu\nabla_\mu^\star\nabla_\mu + M_{\rm cr}(r)\right)
+ m_q\exp(i\omega\gamma_5\tau_3) \right]
\psi(x) 
\eeqa
in the so-called tm-QCD basis if one performs the field redefinition 
\beqa\label{basischange}
\psi_{\rm ph}&=&\exp(i\frac{\omega}{2}\gamma_5\tau_3)\psi, \quad
\bar\psi_{\rm ph}=\bar\psi\exp(i\frac{\omega}{2}\gamma_5\tau_3) .
\eeqa
Here $M_{\rm cr}(r)$ denotes the critical quark mass and $m_q$ is the subtracted 
quark mass defined by $ m_q = m_0 - M_{\rm cr}(r)$ with the bare quark mass $m_0$. Setting the twist angle $\omega$ to zero the critical mass cancels and the standard Wilson mass term remains. This is no longer true for non-zero twist. 

A particular definition for $M_{\rm cr}(r)$ is not relevant for the following argument. However, a crucial assumption is that the critical mass $M_{\rm cr}(r)$ is an odd function of the Wilson parameter $r$,
\beqa\label{SymmetryMcr}
M_{\rm cr}(-r) &=&-M_{\rm cr}(r).
\eeqa
Provided that this is true, one can show that any observable
$\langle O \rangle (r, m_q,\omega)$\footnote{$\langle O \rangle (r, m_q,\omega)$  denotes the expectation value of a local and gauge invariant operator, where the dependence on $r$, $m_{q}$ and $\omega$ is made explicit.}  
 can be O(a) improved by either taking the 
Wilson Average (WA), defined as
\beqa\label{DefWA}
\langle O \rangle^{WA} (r, m_q,\omega)&\equiv&
\frac{1}{2}\left[
\langle O \rangle (r,m_q,\omega) +
\langle O \rangle (-r,m_q,\omega) \right],
\eeqa
or by taking the Mass Average (MA)
\beqa
\langle O \rangle^{MA} (r, m_q,\omega)&\equiv&
\frac{1}{2}\left[
\langle O \rangle (r,m_q,\omega) + (-1)^{P_{{\cal R}_{5}}[O]}
\langle O \rangle (r,-m_q,\omega) \right].
\eeqa
The factor $(-1)^{P_{{\cal R}_{5}}[O]}$ is called the ${\cal R}_{5}$-parity of the operator $O$ \cite{Frezzotti:2003ni}. 
O(a) improvement means that
\beqa
\langle O \rangle^{WA}  (r, m_q,\omega)&=&
\langle O \rangle^{\rm cont} (m_q) + {\cal O}(a^2),
\label{eq:WA} \\
\langle O \rangle^{MA} (r, m_q,\omega) &=&
\langle O \rangle^{\rm cont} (m_q) + {\cal O}(a^2) .
\label{eq:MA}
\eeqa
Using this  one can show that any observable even in $\omega$ is 
automatically O(a) improved at $\omega =\pm\pi/2$ as follows. Consider the Twist Average (TA) at $\omega=\pm\pi/2$:
\beqa
\langle O \rangle^{TA} (r, m_q,\omega=\frac{\pi}{2})&\equiv&
\frac{1}{2}\left[
\langle O \rangle (r,m_q,\omega=\frac{\pi}{2}) +
\langle O \rangle (r,m_q,\omega=-\frac{\pi}{2}) \right].
\eeqa
The expectation value at $\omega = -\pi/2$ on the right hand side is equal to the expectation value at $\omega=2\pi-\pi/2=\pi+\pi/2$, and the mass term in the action, eq.(\ref{eq:twQCD}),
becomes 
\beqa
&&
\bar\psi_{\rm ph}(x)\left[
\left( -a\frac{r}{2}\sum_\mu\nabla_\mu^\star\nabla_\mu + M_{\rm cr}(r)\right)
\exp(-i\left(\pi + \frac{\pi}{2}\right)\gamma_5\tau_3) + m_q\right]
\psi_{\rm ph}(x)\nn \\
&=&
\bar\psi_{\rm ph}(x)\left[
\left( -a\frac{-r}{2}\sum_\mu\nabla_\mu^\star\nabla_\mu + M_{\rm cr}(-r)\right)
\exp(-i\frac{\pi}{2}\gamma_5\tau_3) + m_q\right]
\psi_{\rm ph}(x) ,
\label{eq:change}
\eeqa
provided that  $-M_{\rm cr}(r)=M_{\rm cr}(-r)$. Hence the twist average at $\omega = \pi/2$ is given by 
\beqa
\langle O \rangle^{TA} (r, m_q,\omega=\frac{\pi}{2})&=&
\frac{1}{2}\left[
\langle O \rangle (r,m_q,\omega=\frac{\pi}{2}) +
\langle O \rangle (-r,m_q,\omega=\frac{\pi}{2}) \right],
\label{eq:TA}
\eeqa
which is nothing but the Wilson average in eq.\ \pref{DefWA} and therefore O(a) improved.
If, in addition,  the observable $O$ is even in $\omega$, 
\beqa
\langle O \rangle (r, m_q,\omega=\pm\frac{\pi}{2})&=&
\langle O \rangle^{TA} (r, m_q,\omega=\frac{\pi}{2}),
\label{eq:auto}
\eeqa
the observable $O$ is  automatically O(a) improved without taking  an average. Important examples of $\omega$ even quantities are hadronic masses and some matrix elements \cite{Frezzotti:2003ni}.
%
\subsection{The critical mass}
%
A crucial assumption for the results of the previous subsection is that the critical mass is odd under $r \rightarrow -r$.  This transformation property, however, is not at all obvious. It can be proven in perturbation theory if one defines the critical mass in terms of the pole of the lattice quark propagator \cite{Aoki:1983qi}, but it is likely not to be true non-perturbatively. For example, even if the symmetry properties of the lattice theory imply that the pion mass satisfies \cite{Frezzotti:2003ni}
\be\label{SymmetryOfMpi}
m_{\pi}(r,m_{0}) = m_{\pi}(-r, -m_{0})
\ee
as a function of $r$ and the bare quark mass $m_{0}$, one cannot  conclude that the critical mass, implicitly defined by 
\be\label{DefMcr}
m_{\pi}(r, M_{\rm cr}(r)) = 0,
\ee
is an odd function in $r$. Of course,  property \pref{SymmetryOfMpi} implies  
\be
0\,=\,m_{\pi}(-r, M_{\rm cr}(-r))\,=\, m_{\pi}(r, -M_{\rm cr}(-r)),
\ee
hence both $M_{\rm cr}(r)$ and $-M_{\rm cr}(-r)$ are solutions of \pref{DefMcr}. If equation \pref{DefMcr} has exactly one solution this implies that $M_{\rm cr}(r)$ is indeed an odd function of $r$. However, as soon as \pref{DefMcr} has two or more solutions, this is no longer guaranteed. 

We emphasize that the existence of a massless pion at non-zero lattice spacing  is not trivial since chiral symmetry is explicitly broken by the Wilson term even if the bare mass is set to zero.
A scenario for how a massless pion is realized at non-zero lattice spacing has been proposed a long time ago in Ref.\ \cite{Aoki:1983qi}. The expected phase diagram for Wilson fermions is sketched in figure 1. The solid line represents a second order phase transition line where parity and flavor are spontaneously broken. As a consequence of this spontaneous parity-flavor symmetry breaking the pion mass vanishes along this line. Therefore this phase diagram implies 
the existence of multiple solutions (two for large $g^2$ and ten for small 
$g^2$) to the defining equation \pref{DefMcr} for the critical mass.

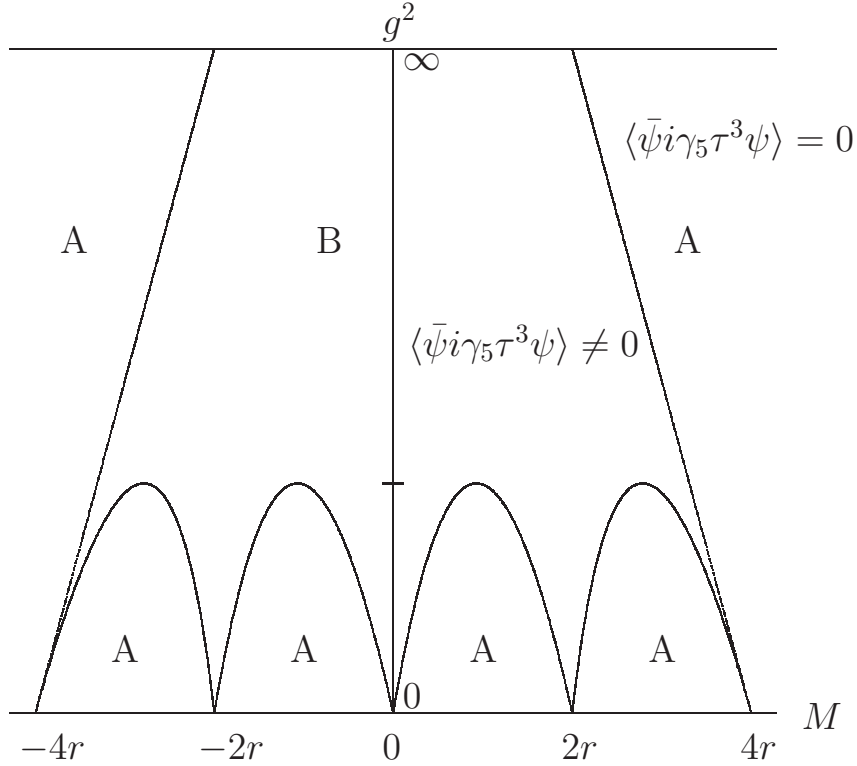
\begin{figure}[t]
\begin{center}
\setlength{\unitlength}{0.68mm}
\begin{picture}(170,150)
\put(10,10){\line(1,0){150}}
\put(85,10){\line(0,1){130}}
\put(10,140){\line(1,0){150}}
\bezier{500}(15,10)(50,140)(50,140)
\bezier{500}(15,10)(40,100)(50,10)
\bezier{500}(50,10)(65,100)(85,10)
\bezier{500}(155,10)(120,140)(120,140)
\bezier{500}(155,10)(130,100)(120,10)
\bezier{500}(120,10)(100,100)(85,10)
\put(165,7){\large $M$}
\put(12,1){\large $-4r$}
\put(47,1){\large $-2r$}
\put(83,1){\large $0$}
\put(118,1){\large $2r$}
\put(153,1){\large $4r$}
\put(83,143){\large $g^2$}
\put(87,136){\large $\infty$}
\put(87,11){\large $0$}
\put(83,55){\line(1,0){4}}
\put(88,80){\large $\langle \bar\psi i\gamma_5\tau^3\psi\rangle \not=0$}
\put(70,100){\large B}
\put(130,120){\large $\langle \bar\psi i\gamma_5\tau^3\psi\rangle =0$}
\put(140,100){\large A}
\put(20,100){\large A}
\put(30,20){\large A}
\put(65,20){\large A}
\put(100,20){\large A}
\put(135,20){\large A}
\end{picture}
\end{center}
\caption{Phase diagram for $N_f=2$ lattice QCD with Wilson fermions,
where $M\equiv m_0 a + 4 r$.
The parity and flavor symmetries are spontaneously broken in phase B.} 
\label{fig:phase}
\end{figure}

Figure 1 is also naturally predicted by WChPT as one of two possible scenarios for the phase diagram for Wilson fermions \cite{Sharpe:1998xm}.\footnote{In the second scenario no massless pion appears for non-zero lattice spacing. We come back to this in section \ref{sectFinalRemarks}.} Moreover, the presence of an $r$-even contribution in $M_{\rm cr}(r)$ at ${\cal O}(a^2)$ has been explicitly shown. This contribution manifests itself as the width of the ``fingers'' in the phase diagram where parity and flavor are spontaneously broken. 
   
In view of these results we assume $M_{\rm cr}(r)$ to have the structure
\beqa\label{McrGeneralStructure}
M_{\rm cr}(r) = M_{\rm odd}(r) + a^2c\, M_{\rm even}(r) \,\equiv\, M^{(1)}_{\rm cr}(r)
\eeqa
where $M_{\rm odd}(r)$ is odd and $M_{\rm even}(r)$ is even under $r \rightarrow -r$. The unknown coefficient $c$ is of mass dimension two and its size is of ${\cal O}(\Lambda_{\rm QCD}^{2})$. Performing the transformation $r\rightarrow -r$ we obtain a second independent solution 
\beqa\label{M2}
-M_{\rm cr}(-r) = M_{\rm odd}(r) - a^2c\,M_{\rm even}(r) \,\equiv\, M^{(2)}_{\rm cr}(r)
\eeqa 
to eq.~\pref{DefMcr}. 
These two solutions correspond to the two critical lines near the physical 
continuum limit, defined at $m_0=0$ (or $M=4r$) and $g^2=0$  in 
Fig.~\ref{fig:phase}, and their distance is of ${\cal O}(a^{2})$.

There exist other definitions for the critical mass than eq.\ \pref{DefMcr}. For example, one can define it in terms of the quark mass entering the PCAC relation. All these definitions differ by terms of ${\cal O}(a)$, and it is again not obvious that a particular definition is odd in $r$. As long as one has not proven this it seems more appropriate to assume the form in eq.\ \pref{McrGeneralStructure} as the general structure for the critical mass. Of course, the details of the functions $M_{\rm odd}(r)$ and $M_{\rm even}(r)$ will differ for each definition of $M_{\rm cr}$.
%
\subsection{Subtleties at $\omega=\pm\pi/2$}
%
Let us assume expression \pref{McrGeneralStructure} for the critical mass and let us see what the consequences are for automatic O(a) improvement at maximal twist. Since the additional contribution in $M_{\rm cr}$ is of ${\cal O}(a^{2}$), the equations (\ref{eq:WA}) and (\ref{eq:MA}) still hold in the presence of  the $M_{\rm even}$ term. However, eq.~(\ref{eq:change}) is modified and now reads 
\beqa
&&
\bar\psi_{\rm ph}(x)\left[
\left( -a\frac{r}{2}\sum_\mu\nabla_\mu^\star\nabla_\mu + M_{\rm cr}(r)\right)
\exp(-i\left(\pi + \frac{\pi}{2}\right)\gamma_5\tau_3) + m_q\right]
\psi_{\rm ph}(x) \nn \\
&=&
\bar\psi_{\rm ph}(x)\left[
\left( -a\frac{-r}{2}\sum_\mu\nabla_\mu^\star\nabla_\mu + M_{\rm cr}(-r)\right)
\exp(-i\frac{\pi}{2}\gamma_5\tau_3)  \right.\nn\\
&&
\hspace{2cm}-2a^2 c\, M_{\rm even}\exp(-i\frac{\pi}{2}\gamma_5\tau_3)+m_q\Bigg]
\psi_{\rm ph}(x) \nn\\
&=&
\bar\psi_{\rm ph}(x)\left[
\left( -a\frac{-r}{2}\sum_\mu\nabla_\mu^\star\nabla_\mu + M_{\rm cr}(-r)\right)
\exp(-i\frac{\pi}{2}\gamma_5\tau_3)  
+m_q^\prime\exp(i\omega^\prime\gamma_5\tau_3)\right]
\psi_{\rm ph}(x), 
\label{eq:change2}
\eeqa
where we have defined
\beqa
m_q^\prime &=& \sqrt{m_q^2+(2a^2c\,M_{\rm even}(r))^2},
\qquad \tan\omega^\prime \,=\, \frac{2a^2c\,M_{\rm even}(r)}{m_q} .
\eeqa
Performing a basis change similar to \pref{basischange} with the angle $\omega'$, eq.~(\ref{eq:TA}) for the twist average gets also modified and is now given by
\beqa
\langle O \rangle^{TA} (r, m_q,\omega=\frac{\pi}{2})&=&
\frac{1}{2}\left[
\langle O \rangle (r,m_q,\omega=\frac{\pi}{2}) +
\langle O \rangle (-r,m_q^\prime,\omega=\frac{\pi}{2}+\omega^\prime) \right]
\label{eq:TA2} .
\eeqa
The twist average is therefore no longer equal to the Wilson average. 
In order to still prove
\beqa
\langle O \rangle^{TA} (r, m_q,\omega=\frac{\pi}{2})&=& \langle O \rangle^{\rm cont} (m_q) + {\cal O}(a^2),
\eeqa
the new mass parameter $m_{q}^{\prime}$ and the angle $\omega^{\prime}$ must satisfy the conditions
\beqa
m_q^\prime &=& m_q +{\cal O}(a^2), \qquad \omega^\prime = {\cal O}(a^2).
\eeqa
The condition for $m_{q}^{\prime}$ is automatically satisfied, but the condition for $\omega^{\prime}$ leads to the inequality
\beqa
 m_q \gg a^{2} \Lambda_{{\rm QCD}}^{3}
\eeqa
in order to guarantee automatic O(a) improvement at $\omega=\pm\pi/2$.
%
\subsection{Alternative definition for $\omega=\pm\pi/2$}\label{AltDef}
%
In this subsection, we propose an alternative definition for the twist angle $\omega$. Setting this angle to 
the values $\pm\pi/2$ results in automatic O(a) improvement for $\omega$-even quantities without any restriction on the size of $m_q$.

We first define $M_{\rm cr}(r)$ as the point where the pion mass vanishes
in the infinite volume limit of lattice QCD at zero twisted mass.
This definition of $M_{\rm cr}(r)$ is unambiguous, contrary
to other definitions such as a vanishing quark mass 
in the axial vector Ward-Takahashi identity, since $M_{\rm cr}(r)$ is equivalent to
a second order phase transition point of the spontaneous parity-flavor breaking
\cite{Aoki:1983qi,Aoki:1986xr,Aoki:1987us,Sharpe:1998xm}. 
Secondly, as we already discussed, there exist (at least)  two independent values of $M_{\rm cr}(r)$, given in \pref{McrGeneralStructure} and \pref{M2}, which are related to each other by 
\be
M_{\rm cr}^{(1)}(r)\,=\,M_{\rm cr}(r),\qquad M_{\rm cr}^{(2)}(r)\,=\,-M_{\rm cr}(-r)\,.
\ee
Neither of these solutions is odd in $r$. 
However, we can define
\beqa
\ovl{M}_{\rm cr}(r) &=&\frac{M_{\rm cr}(r)-M_{\rm cr}(-r)}{2}=-\ovl{M}_{\rm cr}(-r), \\
\Delta M_{\rm cr}(r) &=&\frac{M_{\rm cr}(r)+M_{\rm cr}(-r)}{2}=\Delta M_{\rm cr}(-r),
\eeqa
and $\ovl{M}_{\rm cr}(r) $ is by construction odd in $r$. In terms of $\ovl{M}_{\rm cr}$ and $\Delta M_{\rm cr}$ we now propose an alternative definition for the twist angle $\omega$, which in the physical basis reads
\beqa
\bar\psi_{\rm ph}(x)\left[
\left( -a\frac{r}{2}\sum_\mu\nabla_\mu^\star\nabla_\mu + 
\ovl{M}_{\rm cr}(r) \right)
\exp(-iw\gamma_5\tau_3) + m_q+\Delta M_{\rm cr}(r)\right]
\psi_{\rm ph}(x). 
\label{eq:twQCD2}
\eeqa
With this definition the Wilson average  WA  is the average over 
($m_0=M_{\rm cr}^{(1)}(r)+m_q$, $r$) and 
($m_0=-M_{\rm cr}^{(2)}(r) + m_q$, $-r$), 
which seems to be the natural choice in the presence of two values for 
$ M_{\rm cr}$. Similarly, the mass average MA is the average over 
$m_0=M_{\rm cr}^{(1)}(r) + m_q$ and $m_0= M_{\rm cr}^{(2)}(r) - m_q$ with $r$
fixed. 
Both averages are O(a) improved, since $\ovl{M}_{\rm cr}(r)$ is odd under a 
sign flip in $r$, which is the crucial property for showing O(a) improvement.

At $\omega = -\pi/2 = \pi + \pi/2$ one can show
\beqa
&&
\bar\psi_{\rm ph}(x)\left[
-\left( -a\frac{r}{2}\sum_\mu\nabla_\mu^\star\nabla_\mu + 
\ovl{M}_{\rm cr}(r)\right)
\exp(-i\frac{\pi}{2}\gamma_5\tau_3) + m_q+\Delta M_{\rm cr}(r)\right]
\psi_{\rm ph}(x) \nn \\
&=&
\bar\psi_{\rm ph}(x)\left[
\left( -a\frac{-r}{2}\sum_\mu\nabla_\mu^\star\nabla_\mu + 
\ovl{M}_{\rm cr}(-r)\right)
\exp(-i\frac{\pi}{2}\gamma_5\tau_3) + m_q+\Delta M_{\rm cr}(-r)\right]
\psi_{\rm ph}(x) . \nn
\eeqa
This proves that the twist average at $\omega = \pm \pi/2$  is indeed equal to the Wilson average without any restrictions on the size of $m_q$. 

Let us discuss the meaning of the definition in eq.\ \pref{eq:twQCD2}. For $\omega=0$ and bare mass values  $ M_{\rm cr}^{(2)}(r) < m_0 < M_{\rm cr}^{(1)}(r)$ parity and flavor are  spontaneously broken and the condensate 
$\langle \bar\psi i\gamma_5\tau_3\psi \rangle$ is non-zero.
There is a second order phase transition at the critical values $ M_{\rm cr}^{(1)}(r)$ and $M_{\rm cr}^{(2)}(r)$ where the pion mass vanishes. If one turns on a twisted mass by choosing $\omega \neq 0$, this phase transition becomes a crossover and no massless pion appears. The point $ \ovl{M}_{\rm cr}(r)$ is the center of the parity-flavor broken phase, where the parity-flavor breaking is maximal, i.e.\ $\vert \langle \bar\psi i\gamma_5\tau_3\psi\rangle \vert$ takes its largest value. In this sense a twisted mass term with $\omega=\pm\pi/2$ corresponds to  maximal  twist.

With the definition of $\omega =\pm\pi/2$ in Ref.\cite{Frezzotti:2003ni}
via $M_{\rm cr}^{(1)}(r)$, the massless limit of a maximally twisted quark mass coincides with
the massless limit of an untwisted quark mass for conventional Wilson fermions.
There exists no argument that massless Wilson fermions are O(a) improved, and it is therefore not surprising that an uncanceled O(a) contribution remains when the maximally twisted mass, defined in Ref.\cite{Frezzotti:2003ni}, is made smaller and smaller. This has also been realized in Ref.\cite{Frezzotti:2003ni}. The authors argued that the spontaneous breaking of chiral symmetry should be dominated by the mass term and not by lattice artifacts if one wants to extract physical information from Greens functions. Consequently they imposed the bound $m_{q}\gg a^{2}\Lambda_{\rm QCD}^{3}$ and argued that the chiral limit should only be approached under this condition. As we have shown here, no such bound needs to be imposed as long as one defines the twist angle appropriately. Automatic O(a) improvement at maximal twist can be achieved without any restriction on the quark mass, and it is irrelevant whether the vacuum state is determined by the mass term or by the lattice artifacts. 

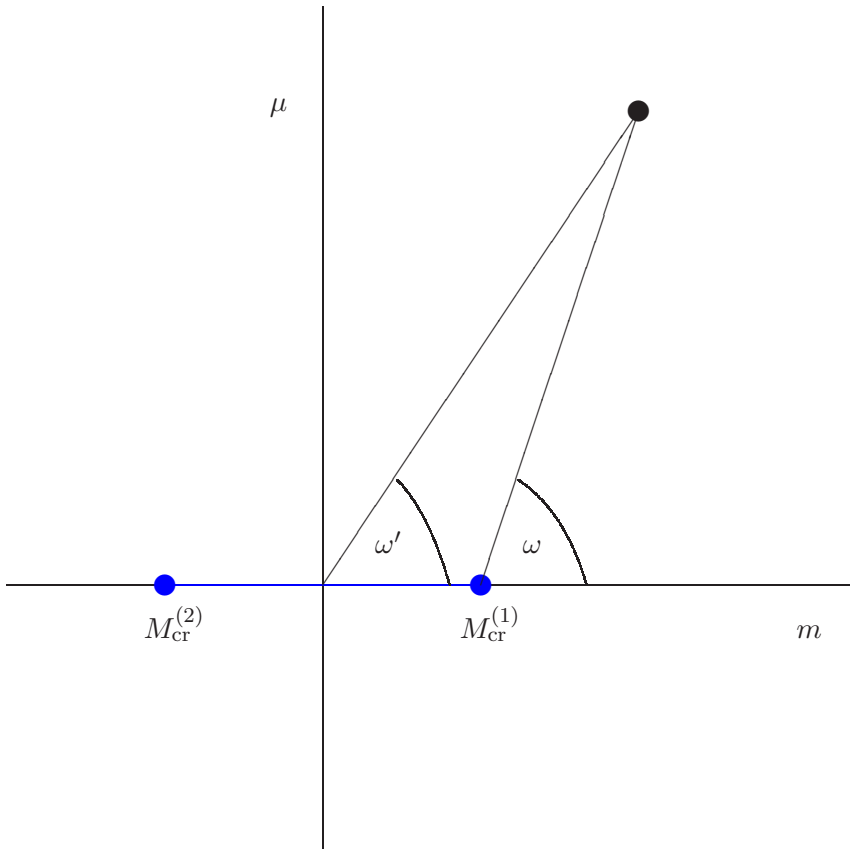
\begin{figure}[t]
\begin{center}
\setlength{\unitlength}{1.4mm}
\begin{picture}(100,80)

\put(20,30){\line(1,0){80}}
\put(50,5){\line(0,1){80}}

\put(35,30){\color{blue}\circle*{2}}
\put(65,30){\color{blue}\circle*{2}}
\put(35,30){\color{blue}\line(1,0){30}}

\put(50,30){\line(2,3){30}}
\put(65,30){\line(1,3){15}}

\put(80,75){\circle*{2}}

\put(55,33){$\omega^{\prime}$}
\put(69,33){$\omega$}
\put(95,25){$m$}
\put(45,75){$\mu$}

\put(33,25){$M^{(2)}_{\rm cr}$}
\put(63,25){$M^{(1)}_{\rm cr}$}

\bezier{200}(62,30)(60,37)(57,40)
\bezier{200}(75,30)(73,37)(68.5,40)

\end{picture}
\caption{\label{fig_2} \small Two different definitions for the twist angle. The angle $\omega$ has been defined in Ref.\cite{Frezzotti:2003ni}, the angle $\omega^{\prime}$ corresponds to the definition in eq.\  \pref{eq:twQCD2}.
}
\end{center}
\end{figure}

The two different definitions of the twist angle are sketched in figure \ref{fig_2}. The angles are approximately equal as long as either $m$ or $\mu$ is much larger than $\ovl{M}_{\rm cr}(r)$. Note that for constant $\omega =\pi/2$ the angle $\omega^{\prime}$ goes  to zero with $\mu\rightarrow 0$.

We finally note that
the $m_q\rightarrow 0$ limit should be taken after the infinite volume limit. This is the usual requirement
in the case that a global symmetry is (expected to be) spontaneously broken.
In practice  one should extrapolate the results calculated at non-zero $m_q$ to the massless point in sufficiently large volume.
%
\section{Wilson Chiral Perturbation Theory (WChPT)}
%
In this section we study the question of automatic O(a) improvement in the chiral effective theory of tmLQCD, i.e.\ Wilson Chiral Perturbation Theory. As an example we compute the tree level pion mass including the lattice spacing effects through ${\cal O}(a^{2})$ for various definitions of the twist angle, and we explicitly show under what conditions the leading lattice spacing effects of ${\cal O}(a)$ cancel at maximal twist.
%
\subsection{Chiral effective Lagrangian}
%
The chiral effective Lagrangian for low-energy tmLQCD has been constructed in Refs.\ \cite{Sharpe:2004ps,Sharpe:2004bv}, where it has been used to analyze the phase diagram of tmLQCD as a function of the quark mass and the lattice spacing (see also Refs.\ \cite{Munster:2004am,Munster:2003ba,Scorzato:2004da} for similar results on the phase diagram). In terms of the $SU(2)$ matrix-valued field $\Sigma$, which transforms under chiral transformations as $\Sigma\rightarrow L\Sigma R^{\dagger}$,  the chiral Lagrangian in Ref.\ \cite{Sharpe:2004ps} reads
\begin{align} \label{ChiralLag}
\mathcal{L}_\chi &= 
 \frac{f^2}{4} \langle\partial_\mu \Sigma \partial_\mu \Sigma^\dagger\rangle
-\frac{f^2}{4} \langle\hat{m}^{\dagger} \Sigma + \Sigma^\dagger\hat{m}\rangle 
-\frac{f^2}{4} \langle\hat{a}^{\dagger} \Sigma + 
               \Sigma^\dagger\hat{a}\rangle \notag \\ 
&\quad
- L_1 \langle\partial_\mu \Sigma \partial_\mu \Sigma^\dagger\rangle^2
- L_2 \langle\partial_\mu \Sigma \partial_\nu \Sigma^\dagger\rangle
      \langle\partial_\mu \Sigma \partial_\nu \Sigma^\dagger\rangle \notag \\
&\quad 
+ (L_4 + L_5/2)\langle\partial_\mu \Sigma^\dagger \partial_\mu \Sigma\rangle
      \langle\hat{m}^{\dagger} \Sigma +  \Sigma^\dagger\hat{m}\rangle
\notag \\ &\quad
+ (W_4 + W_5/2)\langle\partial_\mu \Sigma^\dagger \partial_\mu \Sigma\rangle
      \langle\hat{a}^{\dagger} \Sigma + \Sigma^{\dagger}\hat{a}\rangle\notag \\ &\quad
- (L_6 + L_8/2)  
\langle\hat{m}^{\dagger} \Sigma + \Sigma^\dagger\hat{m}\rangle^2\notag \\ &\quad
- (W_6 + W_8/2)\langle\hat{m}^{\dagger} \Sigma + \Sigma^\dagger\hat{m}\rangle 
      \langle\hat{a}^{\dagger} \Sigma + \Sigma^{\dagger}\hat{a}\rangle
      \notag \\
&\quad
- (W_6'+W_8'/2) \langle\hat{a}^{\dagger} \Sigma +
       \Sigma^{\dagger}\hat{a}\rangle^2.
\end{align}
The angled brackets denote traces over the flavor indices and the short-hand notation 
\bea\label{DefM}
\hat{m} =  2 B m_{R}e^{i\omega\tau_{3}}\equiv 2B(m + i  \mu\tau_3). 
\qquad\hat{a }= 2 W_0 \, a \,,
\eea
is used \cite{Bar:2002nr}. Here $m_{R}, \,\omega$ and $a$ denote the (renormalized) quark mass, twist angle and lattice spacing. The coefficients $B$ and $W_{0}$ are unknown low-energy parameters of dimension one and three, respectively, and $f$ is the pion decay constant in the chiral limit. The $L_{i}$'s are the usual Gasser-Leutwyler coefficients of continuum chiral perturbation theory \cite{Gasser:1983yg,Gasser:1984gg}, while the $W_{i}$'s and $W_{i}^{\prime}$'s are additional low-energy parameters associated with the non-zero lattice spacing contributions \cite{WCPT1,Bar:2003mh}.  

The chiral Lagrangian in Ref.\  \cite{Sharpe:2004ps} contains some more terms than \pref{ChiralLag} since it includes external sources for vector and axial-vector currents as well as for  scalar and pseudo-scalar densities. We do not need these terms in the following and have set them to zero. 

In the Lagrangian \pref{ChiralLag} the twist angle is associated with the mass term. Performing the transformation
\be
\Sigma \,\rightarrow e^{-i\frac{\omega}{2}\tau_{3}} \Sigma e^{-i\frac{\omega}{2}\tau_{3}} \,, 
\ee
the twist angle can be shuffled to the lattice spacing. The Lagrangian is the same as in eq.\ \pref{ChiralLag}, but now parameterized in terms of 
\bea
\hat{m} =  2 B m_{R}, 
\qquad\hat{a }= 2 W_0 \, ae^{-i\omega\tau_{3}} \,.
\eea
%
\subsection{Gap equation}
%
In this section we derive a gap equation for the ground state of the chiral effective theory.  For our purposes it will be enough to only consider the terms of ${\cal O}(m,a,a^{2})$ in the potential energy, which are given by
\beqa\label{PotentialEnergy}
V_{\chi} &=&  \frac{f^2}{4} \langle\hat{m}^{\dagger} \Sigma + \Sigma^\dagger\hat{m}\rangle 
+\frac{f^2 }{4} \langle\hat{a}^{\dagger} \Sigma + 
               \Sigma^\dagger\hat{a}\rangle  + (W_6'+W_8'/2) \langle\hat{a}^{\dagger} \Sigma +
       \Sigma^{\dagger}\hat{a}\rangle^2\,.
\eeqa
We assume the ansatz 
\beqa\label{AnsatzVEV}
\Sigma_0 &=& e^{i \phi \tau_3} 
\eeqa
for the vacuum expectation value (VEV) of the field $\Sigma$. In general, the ground state configuration could have a contribution pointing into a direction orthogonal to $\tau_3$. However, as has already been shown in Refs.\ \cite{Scorzato:2004da,Sharpe:2004ps}, this is not realized for the potential \pref{PotentialEnergy}. 
With this ansatz the potential energy becomes
\beqa
V_{{\chi}}&=& f^2 2B m_R\cos (\phi-\omega) +  f^2 2W_{0} a \cos \phi 
-  f^2 c_2 a^{2}\cos^2\phi,
\eeqa
where we introduced the short-hand notation\footnote{Note that our definition for $c_{2}$ differs by a factor of $f^{2}a^{2}$ from the one in Ref.\  \cite{Sharpe:2004ps}. Furthermore, we have dropped the terms proportional to the quark mass.}
\bea
c_{2} & = & -32\, (2W^{\prime}_{6}+W_{8}^{\prime}) \frac{W_{0}^{2}}{f^{2}}\,.
\eea
In the following we always assume this parameter to be positive, since this sign corresponds to the scenario with spontaneous parity-flavor breaking \cite{Sharpe:1998xm}. The ground state is determined by the {\em gap equation}. 
\beqa\label{originalGapEq}
\frac{d V_{\chi}}{d\phi}&=& f^2 2B m_R\sin(\phi-\omega)+ f^2 2W_{0} a\sin\phi
-2f^2 c_2 a^{2}\sin\phi\cos\phi = 0 ,
\eeqa
which can be rewritten in terms of $m$ and $\mu$ (defined in eq.\ \pref{DefM}) as 
\beqa\label{GapEq_final}
2B \mu\cos\phi &=&\sin\phi\left( 2B m + 2W_{0} a -2c_2 a^{2}\cos\phi\right) .
\eeqa
This equation is invariant under the sign reversal $\mu \rightarrow -\mu,\, \phi \rightarrow -\phi$. This implies that once we have found a solution for positive values of $\mu$ we have also found the solution for negative twisted mass values. Hence, without loss of generality 
we can assume $\mu$ to be positive and we can 
take the square of  equation \pref{GapEq_final}. Setting 
\bea
t& =& \cos\phi
\eea the squared gap equation can be brought into the form
\beqa\label{SquaredGapEQ}
\alpha^{2} t^2 &=& (\chi-t)^2(1-t^2),
\label{eq:gap}
\eeqa
where we introduced
\beqa
\alpha &=& \frac{2B\mu}{2 c_2a^{2}}, \quad 
\chi =\frac{ 2 B m + 2 W_{0}a }{2c_2a^{2}} .
\eeqa
We give some approximate solutions to the gap equation in section \ref{VEV}, but some general statements about the solutions can be made just from the structure of  eq.\ \pref{eq:gap}.  As long as $\chi$ is larger than 1 the gap equation has always two solutions,  one positive and one negative one. Only for $|\chi| < 1$ it can have up to four solutions. If $\alpha \neq 0$ the modulus of the solution is strictly smaller than 1 and $t$ goes to zero for $\alpha \rightarrow \infty$. Finally, $t=0$ is a solution only if $\chi=0$. 
%
\subsection{Pion mass formulae}
%
In order to calculate the pion masses  we expand $\Sigma$ around the vacuum configuration $\Sigma_{0}$. As usual we parametrize the field $\Sigma$ in terms of the pion fields according to 
\beqa
\Sigma(x) &=& \Sigma_{0} \exp \Big(\sum_{i=1}^{3}{i\pi_i(x) \tau_i/f}\Big).
\eeqa
Using this form in expression \pref{PotentialEnergy} for the potential energy we expand in powers of the pion fields. The contribution quadratic in $\pi$ reads
\beqa
V_{\chi,{\rm quad}} &=& \frac{1}{2}\left[ \left(2B m_R\cos(\phi-\omega)+ 2W_{0}a \cos\phi
-2c_2 a^{2}\cos^2\phi \right)\pi\cdot\pi +2c_2a^{2}\sin^2\phi\  \pi_3^2\right],
\eeqa
and the pion masses are therefore given by 
\beqa
m_{\pi_{a}}^2 &=& 2B m_R\cos(\phi-\omega)+ 2W_{0}a \cos\phi -2c_2a^{2}
\cos^2\phi, \qquad {a}= 1,2, \\
m_{\pi_3}^2 &=& m_{\pi_a}^2+2c_2a^{2}\sin^2\phi.
\eeqa
Multiplying the gap equation \pref{GapEq_final} by $\sin\phi$ one easily finds 
\beqa
2 B m_R \cos (\phi-\omega) + 2W_{0}a\cos\phi - 2c_2 a^{2} \cos^2\phi
&=& \frac{2B m + 2W_{0}a}{t} -2c_2a^{2},
\eeqa
which can be used to rewrite the pion mass formulae as
\bea
m_{\pi_{ a}}^2 &=& \frac{2B m + 2 W_{0}a}{t} -2c_2a^{2},\qquad { a} =1,2, \label{pi_a}\\
m_{\pi_3}^2 &=& m_{\pi_{a}}^2 +2c_2a^{2} (1-t^2) .\label{pi_3}
\eea
As expected, the pions are degenerate only for $t=1$, i.e.\ when $\Sigma_{0}$ is proportional to the identity and the flavor symmetry is  unbroken. The appearance of $t$ in the denominator in eq.\ \pref{pi_a} does not imply  a divergence in the pion mass, since $t=0$ is a solution to the gap equation only if $2B m + 2 W_{0}a=0$. This is evident from the alternative expression
\beqa
m_{\pi_{ a}}^2 &=&\frac{2B\mu}{\sqrt{1-t^2}}, \qquad {a} =1,2 ,\label{AlternativeMpi}
\eeqa
for the pion mass, which is easily obtained by rewriting eq.\ \pref{eq:gap} as
\bea
\frac{\alpha^{2}}{1-t^{2}}&=& \left(\frac{\chi}{t}-1\right)^{2}.
\eea
This form is valid as long as $t^2\not= 1$.

Note that the pion mass in the untwisted case does not vanish for $m=0$.
Even though the definition of the quark mass $m$ includes the subtraction of the additive renormalization proportional to $1/a$, it does not include the full subtraction of the critical quark mass \cite{Sharpe:1998xm}. Note in particular the contribution $2c_2a^{2}$ to the critical quark mass, which corresponds to the $r$-even contribution in our ansatz eq.\ \pref{McrGeneralStructure}. 
%
\subsection{VEV and pion masses}\label{VEV} 
%
In this section we present some approximate solutions to the gap equation \pref{eq:gap}. Approximate solutions will be sufficient for our purposes, since we are mainly interested in generic cases where the quark mass $m$ and/or the twisted mass $\mu$ is either much larger or much smaller than the lattice artifacts. In these cases the gap equation usually simplifies and an approximate solution is easily found. 

Consider, for example, the case where both $2B m+2W_{0}a$ and $2B\mu$ are much larger than $2c_{2}a^{2}$, i.e. $\chi\gg1$ and $\alpha\gg1$. In this case we can approximate the gap equation by 
\be
\alpha^{2}t_{0}^{2}\,=\,\chi^{2}(1-t_{0}^{2}),
\ee
and the approximate solution $t_{0}\approx t$ is readily found to be
\be
t_{0}\,=\, \frac{\chi}{\sqrt{\chi^{2} + \alpha^{2}}}\,.
\ee
Once we have found the dominant part $t_{0}$ of the solution, we write $t = t_{0} -\delta$ and substitute this into the gap equation. Since the correction satisfies $\delta \ll t_{0}$ we only keep the terms linear in $\delta$, and the resulting equation is easily inverted to give $\delta$. The result for $t$ can then be used in eqn.\ \pref{pi_a} and \pref{pi_3} in order to obtain approximate expressions for the pion masses. Using this procedure we find the following approximate solutions for the gap equation:
\begin{enumerate}
\item $\chi \gg 1$ and $\alpha \gg 1$ (i.e.\ $2B m+2W_{0}a \ge {\cal O}(a)$ and $2B\mu \ge {\cal O}(a)$). We set $t= t_0-\delta$ and find
\bea
t_0 \,=\, \frac{\chi}{\sqrt{\chi^{2} + \alpha^{2}}},\qquad
\delta \,=\,\frac{\alpha^{2}\chi}
{(\chi^2+\alpha^2)^2}.
\eeqa
In this case the pion masses become
\beqa
m_{\pi_a}^2&=&2c_2 a^2\left[ \sqrt{\chi^2+\alpha^2} - \frac{\chi^2}{\chi^2+\alpha^2}\right],\\
\Delta m_\pi^2 &\equiv&m_{\pi_3}^2-m_{\pi_a}^2 = 2c_2a^{2}
\frac{\alpha^2}{\chi^2+\alpha^2}.
\eeqa
\item $\chi \approx 1 $ and $\alpha \gg 1$ (i.e.\ $2B m+2W_{0}a = {\cal O}(a^2)$ and $2B\mu \ge {\cal O}(a)$). Again, $t = t_{0} - \delta$,
\beqa
t_{0}&=&\frac{\chi}{\alpha},\qquad \delta \,=\,\frac{\chi}{\alpha^{2}},\\
m_{\pi_a}^2&=&2 c_2 a^{2}\left[\frac{\alpha}{1-1/\alpha}-1\right]
\simeq 2 c_2 a^{2} \alpha + {\cal O}\left(\frac{a^{2}}{\alpha}\right)\\
\Delta m_\pi^2 &=& 2c_2a^{2}\left[1-\frac{\chi^2}{\alpha^2}\right]
\simeq 2 c_2a^{2} .
\eeqa
\item  $\chi > 1$ and $\chi \gg \alpha$ (i.e.\ $2B m+2W_{0}a\gg 2B\mu$). The solution of the gap equation is close to one in this case. We define $t = 1- \delta$ and find
\beqa
2\delta & =&\frac{\alpha^2}{(\chi-1)^2+\alpha^2},
\\
m_{\pi_a}^2&=&2c_2a^2\left[\chi-1+\chi\delta\right], \\
\Delta m_\pi^2 &=& 4 c_2 a^{2} \delta .
\eeqa
\item $\chi < 1$ and $\chi \gg \alpha$ (i.e.\ $2B m+2W_{0}a)\gg2B\mu$. In this case we define $ t=\chi-\delta$,
\bea
\delta &=& \frac{\alpha\chi}{\sqrt{1-\chi^2}},\\
m_{\pi_a}^2&=& 2c_2a^{2} \frac{\delta}{\chi-\delta}\simeq 2c_2a^2\frac{\delta}{\chi}
=2c_2a^2\frac{\alpha}{\sqrt{1-\chi^2}},\\
\Delta m_\pi^2 &=& 2c_2a^{2} \left[1-\chi^2+2\chi\delta\right] .
\eeqa
\item  $\chi = 1$ and $\chi \gg \alpha$ (i.e.\ $2B m+2W_{0}a=2c_2a^{2}$ and $2B m+2W_{0}a\gg2B\mu$). We define $t=1-\delta$ and find
\bea
 2 \delta& =& \left(2\alpha\right)^{2/3},\\
m_{\pi_a}^2&=&c_2 a^2 \left( 2\alpha\right)^{2/3},\\
\Delta m_\pi^2 &=& 2c_2 a^2 \left( 2\alpha\right)^{2/3}.
\eeqa
\end{enumerate}
%
\subsection{Twist angle from the Ward-Takahashi identities}
%
In the continuum formulation of  twisted mass QCD one can derive the vector and axial-vector Ward-Takahashi (WT) identities \cite{Frezzotti:2000nk}
\beqa\label{WTIdentities}
\partial_\mu V_\mu^a &=& -2\mu \epsilon^{3ab} P^b,\qquad
\partial_\mu A_\mu^a \,=\, 2 m P^a + 2i\mu S^0\delta^{a3},
\eeqa
where the currents and densities are given as
\beqa
V_\mu^a &=&\bar\psi \gamma_\mu t^a \psi, \quad
A_\mu^a =\bar\psi \gamma_\mu \gamma_5 t^a\psi, \quad
S^0 =\bar\psi \psi,\quad
P^a =\bar\psi \gamma_5 t^a\psi .
\eeqa
A twist angle $\omega_{\rm WT}$ can be defined by
\beqa\label{omegaWT}
\tan\omega_{\rm WT}&=& \frac{\langle \partial_\mu V_\mu^2\ P^1\rangle}
{\langle \partial_\mu A_\mu^1\ P^1\rangle}.
\eeqa
Using the WT identities \pref{WTIdentities} one can easily establish $\tan\omega_{\rm WT} = \mu/m$, i.e. the twist angle defined by the WT identities coincides with the one in the action. 

Similarly, a twist angle $\omega_{{\rm WT}}$ can be defined in Lattice tmQCD. Due to the explicit breaking of chiral symmetry, however, the WT identity for the axial vector receives additional contributions proportional to powers of the lattice spacing. These contributions can be made explicit by deriving the WT identities on the basis of the Symanzik action for Lattice tmQCD: The Pauli term at ${\cal O}(a)$, for example, will give rise to contributions linear in $a$ on the right hand side of the axial-vector WT identity in \pref{WTIdentities}. 

The ratio on the right hand side of \pref{omegaWT} can also be computed in the chiral effective theory. 
To do this we first derive the vector and axial-vector WT identities in the effective theory. Vector and axial-vector transformations of the field are defined by
\beqa
\Sigma &\rightarrow& L \Sigma R^\dagger,
\eeqa
where 
\beqa
R&=&e^{i(\theta^a_V+\theta^a_A)\tau_a}, \quad
L=e^{i(\theta^a_V-\theta^a_A)\tau_a}. 
\eeqa
For a vector transformation ($\theta^a_A=0$) we have $L=R$, while a pure axial-vector transformation is defined by $\theta^a_V=0$ and satisfies $L=R^{\dagger}$.
Under an infinitesimal local variation the field transforms as
\beqa
\delta \Sigma &=& i\theta^a_V [ \tau_a, \Sigma]+ i\theta^a_A \{\tau_a, \Sigma\},\\
\delta \Sigma^\dagger &=& i\theta^a_V [ \tau_a, \Sigma^\dagger]
-i\theta^a_A \{\tau_a, \Sigma^\dagger\} ,
\eeqa
and the variation of the kinetic term plus the potential energy \pref{PotentialEnergy} in the Lagrangian is given by
\beqa\label{deltaL}
\delta L &=&
i\theta^a_V \left[ -\partial_\mu V_\mu^a + X_V^a\right]
+i\theta^a_A \left[ -\partial_\mu A_\mu^a + X_A^a\right], 
\eeqa
where
\beqa
V_\mu^a&=& \frac{f^2}{2}\langle \tau_a \left(\Sigma^\dagger\partial_\mu\Sigma
+\Sigma\partial_\mu\Sigma^\dagger\right)\rangle,\\
A_\mu^a &=&\frac{f^2}{2}\langle \tau_a \left(\Sigma^\dagger\partial_\mu\Sigma
-\Sigma\partial_\mu\Sigma^\dagger\right)\rangle,\\
X_V^a &=& 
-f^2 2\frac{2B\mu}{4}\epsilon^{3ab} P^b, 
\label{eq:WTI1}\\
X_A^a &=& f^2
\frac{2 (2Bm+2W_{0}a)-c_2 a^{2} S^{0}}{4} P^b - if^2 
2\frac{2B\mu}{4} S^0 \delta^{a3}.
\label{eq:WTI2}
\eeqa
In the last two lines we introduced $P^a=\langle \tau_a (\Sigma-\Sigma^\dagger)$ and 
$S^0=\langle \Sigma +\Sigma^\dagger\rangle $.
The variation $\delta L$  implies the WT identities
\be
\partial_\mu V_\mu^a \,=\,X_V^a,\qquad \partial_\mu A_\mu^a \,=\, X_A^a,
\ee
which are the analogue of eq. \pref{WTIdentities} in the effective theory. Using eqs.(\ref{eq:WTI1}, \ref{eq:WTI2}) we find
\beqa\label{omegaWTChPT}
\tan\omega_{\rm WT} &=& \frac{\langle X_V^2\ P^1\rangle}{\langle X_A^1\ P^1
\rangle}=\frac{2B\mu}{2B m + 2W_{0}a -2 c_2a^{2} \cos\phi}\,=\,  \frac{\alpha}{\chi -t} ,
\eeqa
where we used the expansion
\beqa
\langle S^{0}\ P^1\ P^1 \rangle & = & 4  \cos\phi \langle P^1\ P^1\rangle
+ \langle O(\pi^3) \rangle
\eeqa
and dropped the terms cubic in the pion fields. Setting $a=0$ in \pref{omegaWTChPT} we recover the continuum result $\tan\omega_{\rm WT} = \mu/m$. 
%
\section{Maximal twist and  O($\rm\bf a$) improvement}
%
In this section we study the question of automatic O(a) improvement at maximal twist in the case of the pion mass. Before doing this we want to emphasize that setting the angle $\omega$ in the chiral Lagrangian to $\pi/2$ neither corresponds to maximal twist in the sense of Ref.\cite{Frezzotti:2003ni} nor   in the sense of definition \pref{eq:twQCD2}. The reason is the parameterization of the chiral Lagrangian in terms of a renormalized mass that does not contain the  ${\cal O}(a)$ and ${\cal O}(a^{2})$ contributions to the critical mass. It is, however, not difficult to define the previously discussed twist angles in terms of the parameters in the chiral Lagrangian.
%
\subsection{$\omega =\pi/2$}
%
In order to illustrate the possible subtleties of  O(a) improvement we first consider the case where the twist angle in the chiral Lagrangian is taken to be $\pi/2$. With this choice we have $\mu = m_{R}$,  and we need to discuss the following two cases:
\begin{enumerate}
\item  $2 B\mu = 2 B m_R \ge {\cal O}(a)$. This corresponds to case 1 in section \ref{VEV} and the pion mass is given by
\bea
m_{\pi_a}^2 &=&\sqrt{(2B\mu)^2 +(2W_{0}a)^2} ,\\
\Delta m_\pi^2 &=& 2 c_2 a^{2}\frac{(2B\mu)^2}{(2B\mu)^2+(2W_0a)^2}.
\eea
It is obvious that 
the pion mass is O(a) improved for $2B\mu \gg 2W_{0}a$: $m_{\pi_a}^2 \approx 2 B\mu + {\cal O}(a^{2})$. On the other hand, an O(a) correction to the continuum relation $m_{\pi_a}^2 = 2 B\mu$ is present for $2B\mu \approx 2W_{0}a$.  
\item $ 2B\mu=2Bm_{R}\ll 2W_{0} a$ .
In this case we find 
\beqa
m_{\pi_a}^2 &=& 2W_{0}a -2c_2 a^{2} +\frac{ (2B\mu)^2}{2W_{0}a}, \\
\Delta m_\pi^2 &=& 2c_2 a^{2}\frac{(2B\mu)^2}{(2W_{0}a)^2}\ll {\cal O}(a^2). 
\eeqa
Therefore the O(a) term is present in the pion mass. 
In the massless limit the pion mass does not vanish and is instead given by $m_{\pi_a}^2 =  2W_{0}a -2c_2 a^{2}\neq0$. 
\end{enumerate}
Automatic O(a) improvement is guaranteed only for $\mu = {\cal O}(1) = {\cal O}(\Lambda_{\rm QCD})$ if we define maximal twist as $\omega = \pi/2$.

Note that the twist angle defined through the WT identities is given by
\beqa
\tan\omega_{\rm WT} &\simeq&  \frac{2B\mu}{2W_{0}a} \not=\infty.
\eeqa
Therefore $\omega_{\rm WT}=\pi/2+{\cal O}(a)$ for $2B\mu = {\cal O}(1)$ and
$\omega_{\rm WT}\approx 0 $ for $2B\mu \ll 2W_{0}a$.
This result is consistent with the above observation that the pion mass is O(a) improved
only for $\mu =O(1)$.
%
\subsection{Maximal twist of Frezzotti and Rossi}
%
In Ref.\cite{Frezzotti:2003ni} the twist angle  is defined by
\beqa
(m_0-M_{\rm cr}(r))e^{i\omega^\prime \tau_3 \gamma_5}
\eeqa
in lattice QCD. Up to quadratic order in the lattice spacing this corresponds to
\beqa
2B m_R^\prime e^{i\omega^\prime \tau_3} &=&
(2 B m+ 2W_{0} a -2 c_2a^{2}) + i 2 B\mu\,\tau_3
\eeqa
in the effective theory. The angle $\omega^{\prime}$ is related to the parameters in the effective Lagrangian by
\beqa
\tan\omega^\prime &=&\frac{2B\mu}{2 B m + 2W_{0}a -2 c_2a^{2}}
=\frac{2B m_R\sin\omega}{2 B m_R\cos\omega + 2W_{0}a -2 c_2a^{2}}\,,
\eeqa
and $\omega^\prime =\pi/2$ corresponds to $2 B m + 2W_{0}a=2 c_2 a^{2}$.
As before we consider two cases, which correspond to the cases 2 and 5 of subsection \ref{VEV}:
\begin{enumerate}
\item $ 2 B \mu \ge {\cal O}(a)$. 
\beqa
m_{\pi_a}^2 &=& 2 B\mu ,\\
\Delta m_\pi^2 &=& 2 c_2 a^{2}.
\eeqa
The result is O(a) improved  in this case.
The twist angle from the WT identities is given by 
\beqa
\tan \omega_{\rm WT}&\simeq & \frac{2B\mu}{2c_2a^{2}},
\eeqa
so that $\omega_{\rm WT}=\pi/2 +{\cal O}(a^2)$ for $2B \mu = {\cal O}(1)$, 
while $\omega_{\rm WT}=\pi/2 + {\cal O}(a)$ for $2B \mu = {\cal O}(a)$. 
\item $ 2 B\mu \ll 2 c_2a^{2}$.
\beqa
m_{\pi_a}^2 &=&  (c_2a^{2} )^{1/3}(2B\mu)^{2/3},\\
\Delta m_\pi^2 &=& 2(c_2a^{2} )^{1/3}(2B\mu)^{2/3} .
\eeqa
Although all pion masses vanish in the $\mu\rightarrow 0$ limit,
the power $\mu^{2/3}$ is different from the behavior in the continuum limit,
where the pion masses vanish linearly with $\mu$.
The fractional power $2/3$ is the mean-field critical exponent for the
second order phase transition: As the external field $\mu$ decreases,
the correlation length diverges as $\mu^{-1/3}$ at $T=T_c$.

The twist angle from the WT identities becomes
\beqa
\tan \omega_{\rm WT}&\simeq & \left(\frac{2B\mu}{2 c_2a^{2}}\right)^{1/3}.
\eeqa
Therefore $\omega_{\rm WT}\not=\pi/2$. In particular
$\omega_{\rm WT}=0$ at $\mu = 0$.
\end{enumerate}
It seems that automatic O(a) improvement  holds only for 
$2B \mu \ge {\cal O}(a^{2}\Lambda_{\rm QCD}^4)$ if we define maximal twist
by the condition $2Bm +  2W_{0}a-2 c_2a^{2} = 0$.
%
\subsection{New proposal for maximal twist}
%
Finally we consider the alternative definition for the twist angle proposed
in Sect. \ref{AltDef}. In tmLQCD it is defined by
\beqa
\left(m_0-\frac{M_{\rm cr}(r)-M_{\rm cr}(-r)}{2}\right) e^{i\tilde\omega \tau_3},
\eeqa
where $M_{\rm cr}(r)=M_{\rm cr}^{(1)}(r)$ is the critical quark mass as a function of $r$
given in Sect. \ref{AltDef}. This definition corresponds in the effective theory to
\beqa
\tan \tilde\omega &=&\frac{2B\mu}{2Bm + 2W_{0}a}
=\frac{2B m_R\sin\omega}{2Bm_R\cos\omega + 2W_{0}a}.
\label{eq:twistours}
\eeqa
Maximal twist  $\tilde\omega =\pi/2$ therefore implies $2 B m + 2W_{0}a =0$.
In this case $t=\cos \phi = 0$ is the solution of the gap equation, but since the expression \pref{pi_a} for the pion mass is ill-defined in this case, we have to calculate the ratio $(2Bm + 2W_{0}a)/t$ in the
$2Bm + 2W_{0}a\rightarrow 0^+$ limit.

Since $t\ll1$ is the solution for $\chi \ll 1 $, we can solve the approximate gap equation
\beqa
\alpha^{2} t^2 = (\chi-t)^2.
\eeqa
The solution is given by
\beqa
t &=&\frac{2Bm + 2W_{0}a}{2B\mu + 2 c_2a^{2}},
\eeqa
and we therefore obtain
\beqa
m_{\pi_a}^2 &=& 2 B\mu \label{eq:piours},\\
\Delta m_\pi^2 &=& 2 c_2 a^{2},
\eeqa
for the pion masses. This result can also be derived from the  expression \pref{AlternativeMpi} which is valid for all $t^2\not= 1$. 
In this case, O(a) improvement is automatically satisfied, irrespective of the value of $2B\mu$.\footnote{This result is only true for the {\em squared} pion masses. The mass itself of the neutral pion, $m_{\pi_{3}}$, is of ${\cal O}(a)$ for $2B\mu\ll 2c_2a^{2}$. This has similarities to staggered fermions. Even though the staggered fermion action is automatically O(a) improved, the non-Goldstone pion masses are of  ${\cal O}(a)$ for small quark masses 
 \cite{Lee:1999zx,Aubin:2003mg}.}

In addition to the result for the pion masses, the twist angle from the WT identities is calculated as
\beqa
\tan \omega_{\rm WT} &=& \frac{2B\mu+2 c_2a^{2}}{2Bm+2W_{0}a}.  
\eeqa
In the limit  $2Bm+ 2W_{0}a \rightarrow 0^+$ we consistently obtain $\tilde\omega=\omega_{\rm WT}=\pi/2$.

We want to emphasize that our analysis only shows that the leading term linear in the lattice spacing $a$ is absent in the result for the pion mass. There are, of course, subleading terms proportional to $am^{k}$ which must be absent too. In order to show this explicitly one has to consistently include higher order terms in the expression for the potential energy \pref{PotentialEnergy} and the gap equation. This is possible in principle. In practice, however, even the discussion of the O(am) contribution becomes much more involved and goes beyond the scope of this paper.
%
\section{Final remarks}\label{sectFinalRemarks}
%
We pointed out a caveat in the proof for automatic O(a) improvement in tmLQCD at maximal twist if the twist angle is defined as in Ref.\ \cite{Frezzotti:2003ni}. The proof hinges on the fact that the critical quark mass is an odd function of the Wilson parameter $r$. This property, however,  does not hold for the critical quark mass defined as the value where the pion becomes massless, and it is probably not true for other definitions of the critical mass that are currently used. As a result one has to impose the bound $m_{q}\gg a^{2} \Lambda_{\rm QCD}^{3}$ in order to guarantee automatic O(a) improvement at maximal twist. In this paper we gave an alternative definition for the twist angle which does not require such a restriction on the quark mass. Automatic O(a) improvement can be achieved even if the bound $m_{q}\gg a^{2} \Lambda_{\rm QCD}^{3}$ is not satisfied, provided the twist angle is properly defined. 

Having O(a) improvement without a restriction on the quark mass is quite relevant for numerical lattice simulations. Keeping the quark mass large enough such that the inequality $m_{q}\gg a^{2} \Lambda_{\rm QCD}^{3}$ is satisfied would imply fairly large quark and consequently pion masses. To be more explicit  let us assume approximately $300$ MeV for the scale $\Lambda_{\rm QCD}$. In order to satisfy the bound for a lattice spacing of about $0.1$ fm (which is already rather small)  we need to keep the quark mass larger than half the strange quark mass. This would compromise one of the main motivations for using twisted mass  lattice QCD, namely that one can simulate fairly light quark masses. 

Our alternative definition for the twist angle involves the average $\ovl{M}_{\rm cr} =(M^{(1)}_{\rm cr}+M^{(2)}_{\rm cr})/2$ over two values for the critical mass. In practice, however, it might be unnecessary to determine these two values independently in a numerical simulation. The analysis in the chiral effective theory has shown that our definition for the twist angle coincides with the twist angle defined by the WT identities in the case of maximal twist. Tuning the bare untwisted mass parameter in a simulation such that the denominator in \pref{omegaWT}  vanishes realizes therefore maximal twist. 
  
A crucial property for automatic O(a) improvement is eq.\ \pref{SymmetryMcr} which states that the critical mass is an odd function of the Wilson parameter $r$. This cannot be guaranteed if the defining equation \pref{DefMcr} for the critical mass has more than one solution. This is the case if the massless pion is realized by the spontaneous breakdown of parity and flavor \cite{Aoki:1983qi,Aoki:1986xr,Aoki:1987us}. As we already mentioned, it is also possible that equation \pref{DefMcr} has no solution at all and that the pion is never massless at non-zero lattice spacing. This scenario also emerges quite naturally in WChPT as the alternative to the case where parity and flavor are spontaneously broken \cite{Sharpe:1998xm}.  Recent numerical results suggest that
this scenario might be realized when the Wilson plaquette action and the 
unimproved Wilson fermion action is employed 
\cite{Sternbeck:2003gy,Ilgenfritz:2003gw,Farchioni:2004us}. 

Although further confirmation for the existence of 
this scenario is
needed, let us briefly consider this case here.
Without a solution for the defining equation \pref{DefMcr} 
we have to look for a different definition of ${M}_{\rm cr}$. 
A natural choice might be that the pion masses assume their minimal value 
for  $m_{0} = {M}_{\rm cr}$ (in infinite volume). This is an unambiguous 
definition since this point corresponds to a first order phase transition, 
at least in the framework of the chiral effective theory \cite{Sharpe:1998xm,Sharpe:2004ps}. 
Furthermore, the analysis in the chiral effective theory shows that this 
minimum is unique. 
Consequently, this  ${M}_{\rm cr}$ is odd under the sign 
flip in $r$ and the arguments for automatic O(a) improvement in 
Ref.\ \cite{Frezzotti:2003ni} can be applied. 
Our analysis of O(a) improvement for the pion mass in the chiral effective theory can also be performed for this scenario, and the relevant steps are presented in appendix \ref{appendix}.

The numerical results in Ref.\ \cite{Farchioni:2004us} have been obtained with the standard Wilson plaquette action and the unimproved Wilson fermion action.  It is unknown how a change in the lattice action, for example by adding a clover term,  influences the results. A different lattice action can, at least in principle, affect the size and/or sign of the coefficient $c_{2}$ in the chiral Lagrangian of the effective theory, which eventually determines the phase diagram of tmLQCD close to the continuum limit.

Lattice actions with good scaling properties are important for numerical simulations.  Automatic O(a) improvement at maximal twist may give us an extra handle to achieve this. We no longer need to fix the coefficient of the clover term in order to cancel the linear cut-off artifacts. It is an interesting question whether one can tune this coefficient in order to make $c_{2}$ substantially smaller,  i.e.\ to reduce cut-off artifacts at ${\cal O}(a^{2})$. 

%
\section*{Acknowledgments}
%
We would like to thank S.~Sharpe and K.~Jansen for useful discussions. O.\ B.\ would also like to thank M.~Golterman, Y.~Shamir and S.~Sint for discussions during the workshop ``Matching light quarks to hadrons'' in Benasque, where this work was completed. Support of the Benasque Center of Science is gratefully acknowledged. 

This work is supported in part by the Grants-in-Aid for
Scientific Research from the Ministry of Education, 
Culture, Sports, Science and Technology 
(Nos. 13135204, 15204015, 15540251, 16028201). 
O.\ B.\ is supported in part by the University of Tsukuba Research Project.
%
\appendix
%
\section{The $\mathbf c_2 < 0$ case}\label{appendix}
It has been pointed out recently that tmLQCD undergoes a first order phase transition at small $\mu$ for $\beta = 5.2$ \cite{Farchioni:2004us}. The authors interpret the existence of this first order phase transition as the alternative scenario in WChPT where $c_2 < 0$ \cite{Sharpe:1998xm}. 
Motivated by these results we extend our analysis to this case. (A similar analysis has already been made in Refs.\cite{Munster:2003ba,Scorzato:2004da,Sharpe:2004ps}.) 

At $\mu = 0$, the vacuum expectation value has a gap at $\tilde c_1 =
2 B m_R + 2W_{0}a = 0$:
\beqa
\Sigma_0 &=&\left\{
\begin{array}{rl}
1  & \mbox{for }\tilde c_1 > 0 \\
-1 &  \mbox{for }\tilde c_1 < 0 \\
\end{array} \right.  .
\eeqa
Accordingly the pion masses become
\beqa
m_{\pi_a}^2&=& m_{\pi_3}^2 = \vert \tilde c_1\vert -2 c_2a^{2}
\eeqa
and remain massive for $\tilde c_1 = 0$,
\beqa
m_{\pi_a}^2&=& m_{\pi_3}^2 \,=\,  -2 c_2a^{2}\, >\, 0 .
\eeqa

We now consider how this result changes for non-zero $\mu$ by solving approximately the gap equation \pref{eq:gap}.
\begin{enumerate}
\item Small $\mu$  ($ 2B\mu \ll \vert \tilde c_1\vert - 2c_2 a^2$)\\
The first order phase transition persists in this case:
\beqa
t&=\left\{
\begin{array}{rl}
1-\delta & \mbox{for }\tilde c_1 > 0 \\
-1+\delta & \mbox{for }\tilde c_1 < 0 \\
\end{array} \right. ,
\label{eq:smallMu}
\eeqa
where
\beqa
\delta =\frac{1}{2}\left(\frac{ 2B\mu}{\vert \tilde c_1\vert -2 c_2 a^{2}}
\right)^2\, =\,{\cal O}(\mu^2).
\eeqa
The pion masses are given by
\beqa
m_{\pi_a}^2 &=&\vert \tilde c_1\vert(1+\delta)-2 c_2a^{2}\\
\Delta m_\pi^2 &=& 4c_2 a^2 \delta < 0 .
\eeqa
The point where the first order phase transition occurs ($\tilde c_1 = 0$) 
corresponds to maximal twist according to our new proposal, $\tilde \omega = \pm\pi/2$.
\item Large $\mu$ ($2B\mu \gg \vert \tilde c_1\vert - 2c_2 a^2$)\\
For large $\mu$ we can neglect the term proportional to $c_2$ in eq.\ \pref{originalGapEq} and the first order phase transition
disappears: 
\beqa
t\,=\, \frac{\tilde c_1}{\sqrt{\tilde c_1^2 + (2 B\mu)^2}} .
\eeqa
Therefore $\vert t \vert < 1$ and no gap exits at $\tilde c_1 = 0$.
The pion masses are given by
\beqa
m_{\pi_a}^2 &=& \sqrt{(2 B\mu)^2+ \tilde c_1^2} -2c_2a^{2}\\
\Delta m_\pi^2 &=& 2c_2a^{2}\frac{(2B\mu)^2}{(2B\mu+ \tilde c_1^2)^2} < 0 .
\eeqa
\item General $\mu$ for $\tilde c_1=0$\\
In order to estimate the value of $\mu$ at which the first order phase 
transition disappears, we consider the case $\tilde c_1 =0$ for arbitrary values of $\mu$.
In this case, the solution to the gap equation is given by
\beqa
t &=& \left\{
\begin{array}{cll}
\phantom{-}\sqrt{1-\dfrac{(2B\mu)^2}{(2c_2a^{2})^2}},  & \vert 2 B\mu\vert < -2 c_2a^{2},
& \tilde c_1 \rightarrow 0^+ \\
-\sqrt{1-\dfrac{(2B\mu)^2}{(2c_2a^{2})^2}},  & \vert 2 B\mu\vert < -2 c_2a^{2},
&\tilde c_1 \rightarrow 0^- \\
0, & \vert 2 B\mu\vert \ge -2 c_2a^{2}, & \mbox{for all } \tilde c_1\\
\end{array} \right. .
\eeqa 
Therefore there is no first order phase transition
for $\vert 2 B \mu \vert > -2c_2a^{2}$, and $ 2 B\mu = \mp 2c_2a^{2}$ 
are the two endpoints of the first order phase transition line. 
It is easy to verify that the solution, expanded in term of small $\mu$ as
\beqa
t &=& \left\{
\begin{array}{cll}
\phantom{-}1-\frac{1}{2}\dfrac{(2B\mu)^2}{(2 c_2a^{2})^2},  & \vert 2 B\mu\vert < -2 c_2a^{2},
& \tilde c_1 \rightarrow 0^+ \\
-1+\frac{1}{2}\dfrac{(2B\mu)^2}{(2c_2a^{2})^2}, 
& \vert 2 B\mu\vert < -2c_2a^{2}, &
\tilde c_1 \rightarrow 0^- \\
\end{array} \right. ,
\eeqa
agrees with the one in eq.(\ref{eq:smallMu}) for $\tilde c_1=0$. 
The corresponding pion masses are given by
\beqa
m_{\pi_a}^2 &=&\left\{
\begin{array}{ll}
-2a^{2}c_2,  & \vert 2 B\mu\vert < -2 c_2a^{2} \\
\vert 2 B\mu\vert , & \vert 2 B\mu\vert \ge -2 c_2a^{2} \\
\end{array}\,, \right.  \\
\Delta m_\pi^2 &=&\left\{
\begin{array}{ll}
2c_2a^{2}\left(\frac{2B\mu}{2 c_2a^{2}}\right)^2  < 0,
& \vert 2 B\mu\vert < -2 c_2 a^{2}\\
2 c_2a^{2} < 0, & \vert 2 B\mu\vert \ge -2 c_2a^{2} \\
\end{array} \right.  .
\eeqa
Therefore  $m_{\pi_3}^2 = 0$ at the two endpoints of the first order phase transition line,
$\vert 2 B \mu \vert = -2 c_2a^{2}$. According to eq.(\ref{eq:twistours}),
$\tilde c_1 = 0$ corresponds to $\tilde \omega =\pi/2$ ($-\pi/2$) for positive (negative) values of $\mu$.
\end{enumerate}
Note that the mass difference is negative, i.e.\ the charged pions are heavier than the neutral one, in contrast to the $c_{2}>0$ case. Hence, as has already been pointed out in Ref.\ \cite{Scorzato:2004da}, the sign of the coefficient $c_{2}$ can be determined, at least in principle, by measuring the masses of the charged and neutral pions.
%
\bibliography{basename of .bib file}
%

\end{document}